\documentstyle[11pt,emulateapj]{article}
\received{}
\accepted{}
\journalid{}{}
\articleid{}{}

\lefthead{Mart\'\i n et al$.$}  

\righthead{Keck NIRC observations of
           planetary-mass candidate members in $\sigma$ Orionis}

\slugcomment{Accepted for publication in ApJ Letters}

\newcommand{\mjup}{\,$M_{\rm Jup}$}
\newcommand{\msol}{\,$M_{\odot}$}
\newcommand{\sigori}{$\sigma$\,Orionis}

\begin{document}

\title{Keck NIRC Observations of Planetary-mass Candidate 
       Members in the $\sigma$\,Orionis Open Cluster}

\author{E. L. Mart\'{\i}n}  
\affil{Institute of Astronomy, Univ. of Hawaii at Manoa, 2680 Woodlawn
  Drive, Honolulu, HI 96822, U.S.A.  {\em ege@ifa.hawaii.edu}}
     
\author{M. R. Zapatero Osorio } 
\affil{Div. of Geological and Planetary Sciences, California Institute
  of Technology, Pasadena, CA 91125, U.S.A.  {\em
    mosorio@gps.caltech.edu}}

\author{D. Barrado y Navascu\'es}
\affil{Universidad Aut\'onoma de Madrid, E--28049 Madrid, Spain.  {\em
    barrado@pollux.ft.uam.es}}

\author{V.\,J.\,S.\,B\'ejar}
\affil{Instituto de Astrof\'\i{}sica de Canarias, E-38200 La Laguna,
  Tenerife, Spain. {\em vbejar@ll.iac.es}}

\and
\author{R. Rebolo}
\affil{Instituto de Astrof\'\i{}sica de Canarias, E-38200 La Laguna,
  Tenerife, Spain; and Consejo Superior de Investigaciones Cient\'\i
  ficas, Spain.  {\em rrl@ll.iac.es}}

\begin{abstract}
  We present $K$-band photometry and low-resolution near-infrared
  spectroscopy from 1.44 to 2.45\,$\mu$m of isolated planetary-mass
  candidate members in the \sigori~cluster found by Zapatero Osorio et
  al$.$ The new data have been obtained with NIRC at the Keck~I
  telescope.  All of our targets, except for one, are confirmed as
  likely cluster members.  Hence, we also confirm that the
  planetary-mass domain in the cluster is well populated.  Using our
  deep $K$-band images we searched for companions to the targets in
  the separation range 0.3\arcsec~to 10\arcsec~up to a maximum faint
  limit of $K$=19.5\,mag.  One suspected companion seems to be an
  extremely red galaxy.  The near-infrared colors of the
  \sigori~substellar members indicate that dust grains condense and
  settle in their atmospheres.  We estimate that the surface
  temperatures range from 2500\,K down to 1500\,K.  The spectroscopic
  sequence covers the full range of L subclasses, and the faintest
  object is tentatively classified as T0.  These targets provide a
  sequence of substellar objects of known age, distance and
  metallicity, which can be used as benchmark for understanding the
  spectral properties of ultracool dwarfs.
\end{abstract}


\section{Introduction}
Deep photometric surveys of star-forming regions and young open
clusters have revealed a population of young substellar objects
extending from the substellar borderline ($M\,<\,0.075$\msol) to
masses below the deuterium-burning limit (0.013\msol~or $\sim$
14\mjup, where 1\,\msol\,=\,1047\,\mjup). Recent examples include
searches in IC\,348 (Najita, Tiede, \& Carr \cite{najita00}), the
Trapezium (Lucas et al$.$ \cite{lucas01}; Hillenbrand \& Carpenter
\cite{hillenbrand00}; Luhman et al$.$ \cite{luhman00}), and the
\sigori~cluster (Zapatero Osorio et al$.$ \cite{zapatero00}).  These
discoveries are leading to renewed theoretical work in the formation
of substellar objects (Boss \cite{boss01}; Reipurth \& Clarke
\cite{reipurth01}).  The masses estimated for these objects assume
membership in the cluster, and thus known age, distance and
metallicity. This assumption needs to be tested with follow-up
observations. The sequence of faint cluster members provides a
homogeneous sample to study the relation between temperature and
spectral properties.  The atmospheres of substellar objects are
complicated, being difficult to disentangle the effects of gravity,
metallicity and temperature. Dust grains are expected to condense and
become an important source of opacity that affects the colors and
spectra (Allard et al$.$ \cite{allard01}). The Pleiades substellar
sequence has been identified (e.g., Bouvier et al$.$ \cite{bouvier98};
Stauffer et al$.$ \cite{stauffer98}; Zapatero Osorio et al$.$
\cite{zapatero97}) down to the lowest mass known Pleiad, Roque\,25,
with a mass determined at 0.035\,\msol~(Mart\'\i n et al$.$
\cite{martin98}).  The surveys in regions younger than the Pleiades
are sensitive to lower masses because substellar objects contract and
cool quickly with time. Hence, luminosity and mass functions can be
extended to lower masses. This is the case of the \sigori~cluster
(B\'ejar et al$.$ \cite{bejar01}). It is a prime location to search
for substellar objects because of its young age ($\sim$5\,Myr),
distance ($\sim$350\,pc) and low reddening.  Here we present
near-infrared (NIR) photometry and spectroscopy for a sample of
isolated planetary-mass candidate members of the \sigori~cluster.  The
colors of the objects can be explained with models that include the
presence of dust grains.  The spectra provide strong evidence that
these objects are indeed ultracool cluster members.

\section{Observations and results}
Zapatero Osorio et al$.$ (\cite{zapatero00}) reported the discovery of
18 isolated planetary-mass candidate members in the \sigori~cluster.
For two of them, namely S\,Ori\,52 and 56, they presented optical
spectroscopy, and for an additional object they presented NIR
spectroscopy. We have found that the identification of this object was
confused. The NIR spectrum shown in Fig.~3 of Zapatero Osorio et al$.$
corresponds to S\,Ori\,62, not S\,Ori\,60 as stated in their Table~1.
S\,Ori\,62 and 60 have indeed very similar magnitudes, spectral types
and masses. Hence, the conclusions of Zapatero Osorio et al$.$ remain
unchanged.

On 2000 December 16, we observed with the Keck\,I NIR camera NIRC
(Matthews \& Soifer \cite{matthews94}). Meteorological conditions were
photometric and the seeing ranged from 0.3\arcsec~to 0.7\arcsec, as
measured in the NIRC images.  We obtained $K$-band images of eight
planetary-mass candidate members of \sigori, and grism spectroscopy
for four.  The $K$-band images consisted of 12 coadds of 10~s each.
The spectra were obtained with 3 coadds of 100~s each using the $HK$
filter and the grism gr120. This configuration provides simultaneous
coverage of the $H$- and $K$-band spectra (from 1.44 to 2.45\,$\mu$m)
with very low dispersion (5.9\,nm\,pix$^{-1}$).

One object located 1.6\arcsec~west and 3.7\arcsec~north of S\,Ori\,47, a
spectroscopically confirmed cluster member (Zapatero Osorio et al$.$
\cite{zapatero99}), was discovered in NIR images obtained with the
2.2-m telescope in Calar Alto on 2000 February 16--21. We recovered it
in a CCD image of S\,Ori\,47 collected with the Low-Resolution Imaging
Spectrograph (LRIS; Oke et al$.$ \cite{oke95}) at Keck\,II on 1998
December 21, and we were able to measure a magnitude of
$I$=24.2$\pm$0.4.  Thus, this is a very red object, with
$I-K$=5.2$\pm$0.5.  We call it S\,Ori\,J053814.4--024012. Deep images
(total exposure time 360\,s in $K$-band) and grism spectroscopy of
S\,Ori\,J053814.4--024012 were obtained with NIRC.

The data were reduced using standard routines for bias-substraction,
flat-field correction and sky-substraction within the
IRAF\footnote{IRAF is distributed by National Optical Astronomy
  Observatories, which is operated by the Association of Universities
  for Research in Astronomy, Inc., under contract with the National
  Science Foundation.} enviroment.  Telluric absorption was removed
using the NIRC spectrum of the G0 star SAO\,112561, observed at
similar airmass as the program objects.  Photometric calibration was
made using the 2MASS Second Incremental Release Point Source Catalog.
In Table~1 we provide calibrated $K$-band photometry for all the
targets.  The $I$- and $J$-band photometry are taken from Zapatero
Osorio et al$.$ (\cite{zapatero00}).

\begin{figure}
\epsscale{0.6}
\plotone{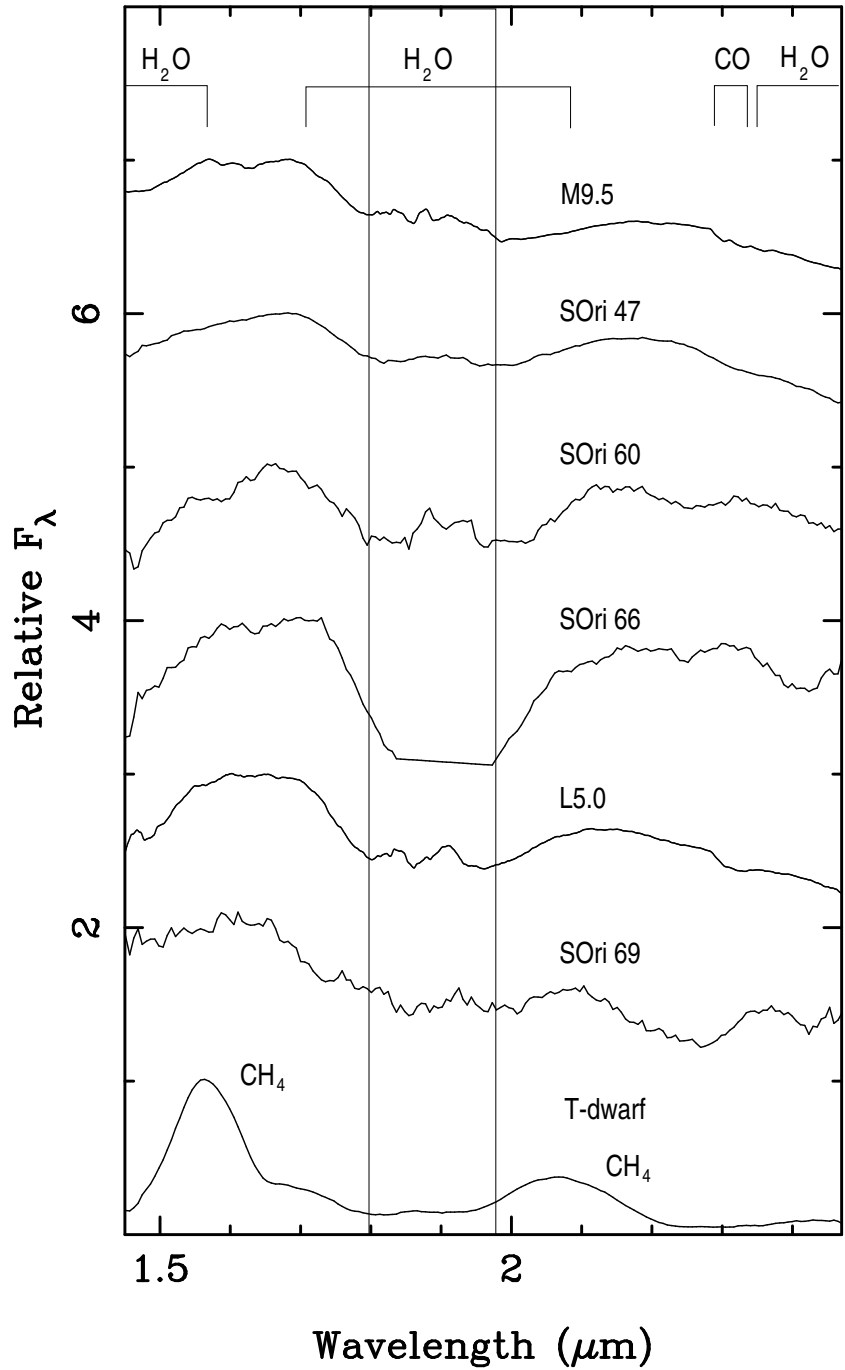}
\caption{\label{fig1} NIRC spectra for four
  program objects in \sigori.  The spectra of field ultracool dwarfs
  of known spectral type are shown for comparison.  The M9.5 dwarf is
  BRI\,0021$-$0214, the L5 dwarf is DENIS-P J1228.2$-$1547AB, and the T
  dwarf is 2MASSI\,J1047539+212423. The first two spectra were taken
  from Leggett et al$.$ (\cite{leggett01}).  The T-dwarf spectrum was
  obtained with NIRC in the same night as the program objects.  A
  boxcar smoothing of 15 points has been applied to all the spectra.
  Vertical lines bracket a spectral region strongly affected by
  telluric absorption.}
\end{figure}

The final smoothed NIRC spectra are shown in Fig.~\ref{fig1}.  Our new
spectrum of S\,Ori\,47 is very similar to that presented by Zapatero
Osorio et al$.$ (\cite{zapatero00}).  S\,Ori\,60 and 66 present broad
strong absorption features that we identify as water bands.
Identification of the CO bands beyond 2.2\,$\mu$m is rendered
difficult by the low spectral resolution and modest S/N ratio of the
data.  S\,Ori\,69 also shows water bands, and possibly the methane
bandhead at 2.2\,$\mu$m.  We do not see, however, any hint of methane
in the $H$-band. Since the spectrum of S\,Ori\,69 is rather noisy, we
do not place high confidence on the detection of methane.
Nevertheless, it is clear that the $K$-band flux is depressed with
respect to the $H$-band in comparison with brighter cluster members.
Allard et al$.$ (\cite{allard01}) have shown that this is expected
when dust grains condense below the photosphere. Barrado y Navascu\'es
et al$.$ (\cite{barrado01}) have obtained optical spectra of
S\,Ori\,60 and 66 with the Very Large Telescope.  They measured
H$\alpha$ emission equivalent widths of 25\,\AA~and 100\,\AA~in
S\,Ori\,60 and 66, respectively.  They estimated spectral types of
L2.0$\pm$0.5 for S\,Ori\,60 and L3.5$\pm$2.0 for S\,Ori\,66 by
comparison with field ultracool dwarfs.  Our data indicates that
S\,Ori\,66 is slightly cooler than S\,Ori\,60 because the water bands
are deeper.  Using the water index in the $H$-band defined by Delfosse
et al$.$ (\cite{delfosse99}) and calibrated with the spectral types of
Mart\'\i n et al$.$ (\cite{martin99}), we estimate spectral types of
L5.5 for S\,Ori\,60 and L6 for S\,Ori\,66.  We tentatively classify
S\,Ori\,69 as a T0 dwarf.  A T0 dwarf shows clear methane absorption
in the $K$-band (bandhead at 2.2\,$\mu$m, which may also be weakly
present in the L5 dwarf DENIS-P\,0205--15; Burgasser et al$.$
\cite{burgasser01}) but not in the $H$-band (bandhead at 1.6\,$\mu$m).
The strong methane band at 3.3\,$\mu$m has been detected in mid-L
dwarfs (Noll et al$.$ \cite{noll00}).
 
In order to compare our program objects with the Trapezium objects
observed by Lucas et al$.$ (\cite{lucas01}), we measured the water
index W defined by those authors.  We found W\,=\,0.66 for S\,Ori\,47,
and W\,=\,0.49 for S\,Ori\,60 and 66.  Our values are within the range
of values measured by Lucas et al$.$ for the coolest Trapezium
objects, and suggest that the Trapezium and \sigori~objects have
similar temperatures. We do not find, however, the triangular shapes
in $H$-band spectra reported by Lucas et al$.$ for some of their
objects.  This could be due to the difference in age between the
Trapezium ($\le$1\,Myr) and the \sigori~cluster ($\sim$5\,Myr), which
would imply that the members of the later cluster have higher
gravities. The dissimilarities could also be due to some other
effects.

\begin{figure}
\epsscale{0.6}
\plotone{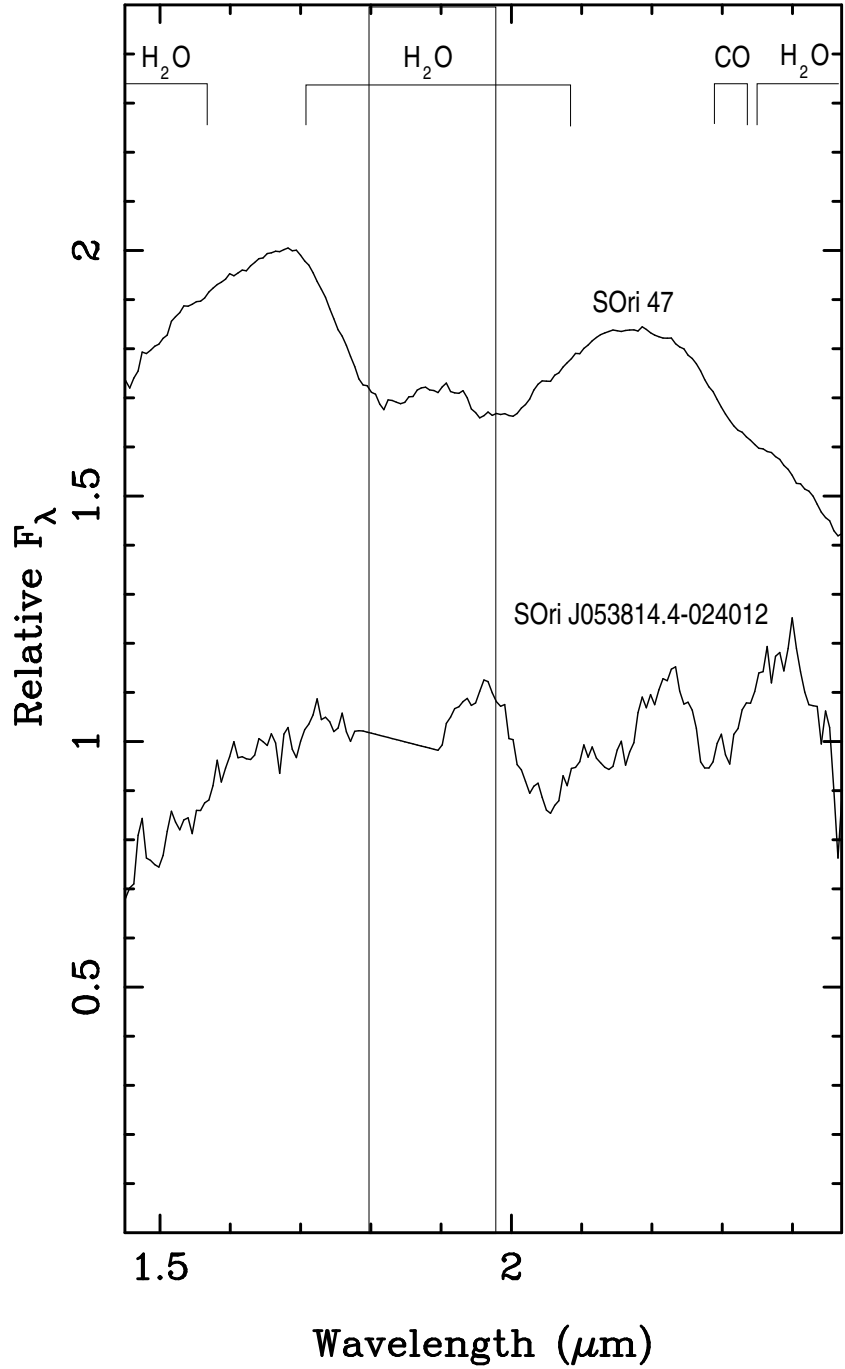}
\caption{\label{fig2} NIRC spectra for S\,Ori\,47
  and S\,Ori\,J053814.4--024012 obtained in the same night. A boxcar
  smoothing of 15 points has been applied to all the spectra.  The two
  vertical lines bracket a spectral region strongly affected by
  telluric absorption.}
\end{figure}

The NIRC spectrum of the possible very low-mass companion to S\,Ori\,47
is shown in Fig.~\ref{fig2}.  We do not find strong methane or water
bands in S\,Ori\,J053814.4--024012, suggesting that it is not a
substellar object.  We discuss this object further in the next
section.

\section{Discussion and final remarks}
All of our targets have very red colors, $I-K>4.5$ and $J-K>1.0$,
except for one of them, namely S\,Ori\,57. This object is probably a
field M dwarf.  Model atmospheres of ultracool dwarfs need to consider
the effects of dust grains for producing such red colors. Allard et
al$.$ (\cite{allard01}) have computed synthetic colors using two extreme
approximations for the treatment of dust.  Dusty models assume that
dust grains form in the upper atmosphere and remain there, producing a
greenhouse effect.  On the other hand, Cond models assume total
gravitational condensation of dust grains below the photosphere, and
effectively all dust opacity is neglected.

\begin{figure}
\epsscale{0.9}
\plotone{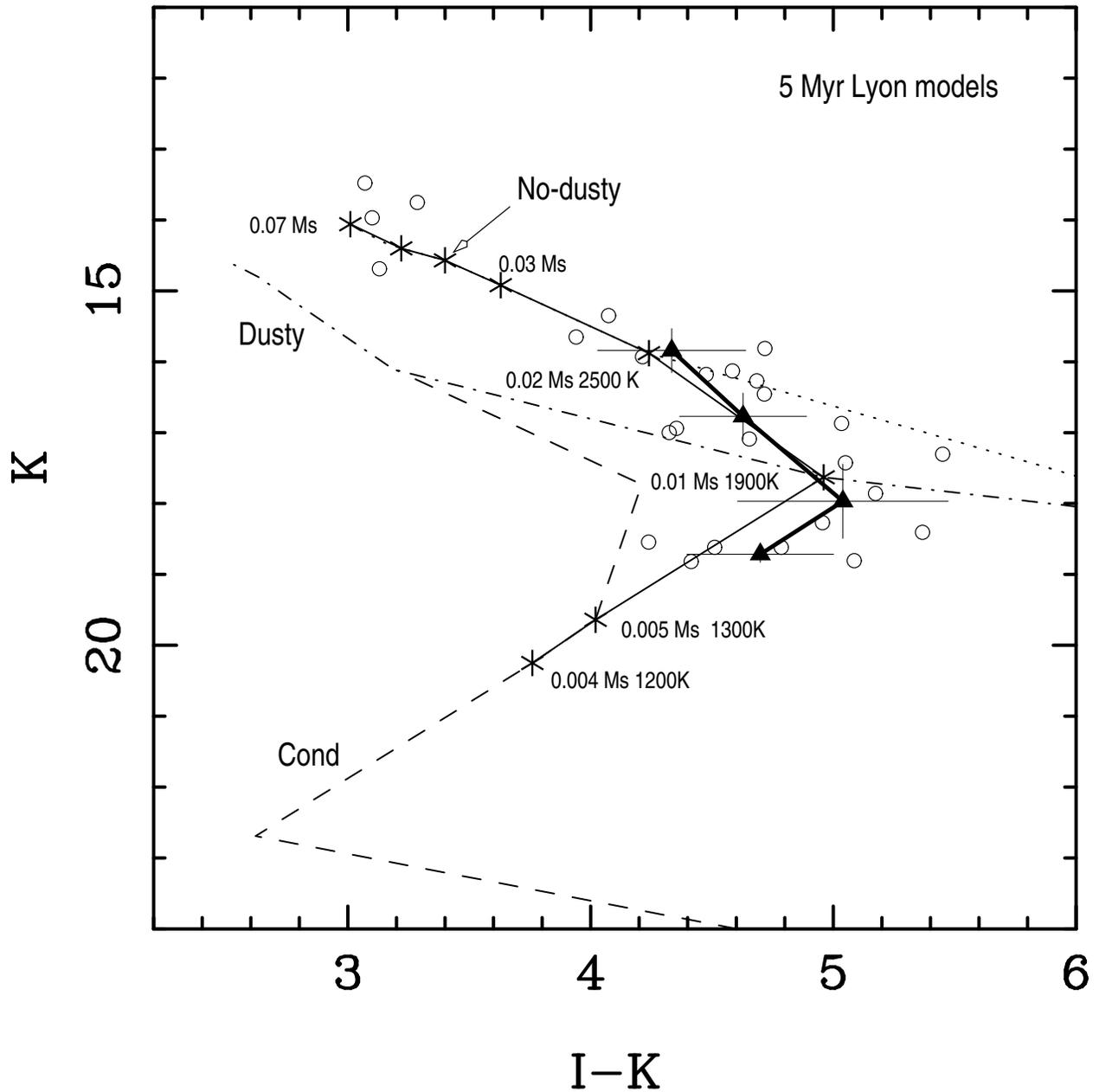}
\caption{\label{fig3} $K$ vs $I$-$K$
  color--magnitude diagram. The 5 Myr isochrones from Chabrier et
  al$.$ (\cite{chabrier00}) are displayed (Nextgen no-dusty
  models---dotted line, Dusty models---dot-dashed line, and Cond
  models -- dashed line).  The thin solid line with asterisks is a
  best-guess isochrone that combines the models.  Empty circles are
  substellar members in \sigori. The thick solid line and filled
  triangles represent the mean locus of the \sigori~objects.}
\end{figure}

In Fig.~\ref{fig3}, we show a color-magnitude diagram with theoretical
isochrones from the models of Chabrier et al$.$ (\cite{chabrier00})
for an age of 5\,Myr (for a detailed discussion of the cluster age and
distance see B\'ejar et al$.$ \cite{bejar01} and Zapatero Osorio et
al$.$, in preparation).  The objects plotted on the figure result from
the combination of our photometry with that of B\'ejar et al$.$
(\cite{bejar01}) and Zapatero Osorio et al$.$ (\cite{zapatero00}).
Likely non-members of the cluster, such as S\,Ori\,J053814.4--024012
and S\,Ori\,57, have been excluded. Three regimes can be distinguished
in this figure: {\em (i)} Dust free: for colors bluer than
$I-K$\,=\,4.4 (corresponding to surface temperatures $T_{\rm
  eff}$\,$>$\,2500\,K, and masses $M$\,$>$\,0.02\,\msol) atmosphere
models without dust grains successfully reproduce the observed
photometric sequence. {\em (ii)} Dusty: the mean locus of objects
deviates from the dust-free isochrone and intersects the dusty
isochrone for $I-K$ colors between 4.4 and 5.0 ($T_{\rm
  eff}$\,$\sim$\,2500--1900\,K, $M$\,$\sim$\,0.02--0.01\,\msol). This
temperature range corresponds to spectral types M9 through L3 (Basri
et al$.$ \cite{basri00}), in agreement with our observations of
S\,Ori\,45 (M8.5, B\'ejar, Zapatero Osorio, \& Rebolo \cite{bejar99})
and S\,Ori\,62 (L4).  S\,Ori\,47 is a benchmark object for this
regime.  {\em (iii)} Settling dust: the dusty isochrone predicts that
cooler objects should be redder but the observed locus of objects
appears to turn blueward in $I-K$ color at the faintest end of the
current study.  The Cond models have much bluer colors than the Dusty
models. Thus, the colors of the faintest cluster members may be
turning blue due to gravitational settling of dust grains.  The
transition from the red L dwarf colors to the blue T dwarf colors
appears to be driven by the gravitational settling of dust grains.
S\,Ori\,66 is a benchmark object for this regime.  The possible
presence of methane absorption in the $K$-band spectrum of S\,Ori\,69
suggests that it could be an early T dwarf with $T_{\rm
  eff}$\,$\sim$\,1500\,K.  The lack of obvious methane absorption in
the $H$-band spectrum, and the red $J-K$ color of this object indicate
that it is warmer than the T-dwarfs reported by Leggett et al$.$
(\cite{leggett00}).  The small difference in magnitude and color
between S\,Ori\,66 and 69 (0.54 mag in $K$) suggests that the gap
between the latest L dwarf (L6) and earliest T dwarf (T0) has been
bridged provided the fact that the presence of methane is confirmed in
S\,Ori\,69.

\begin{figure}
\epsscale{0.9}
\plotone{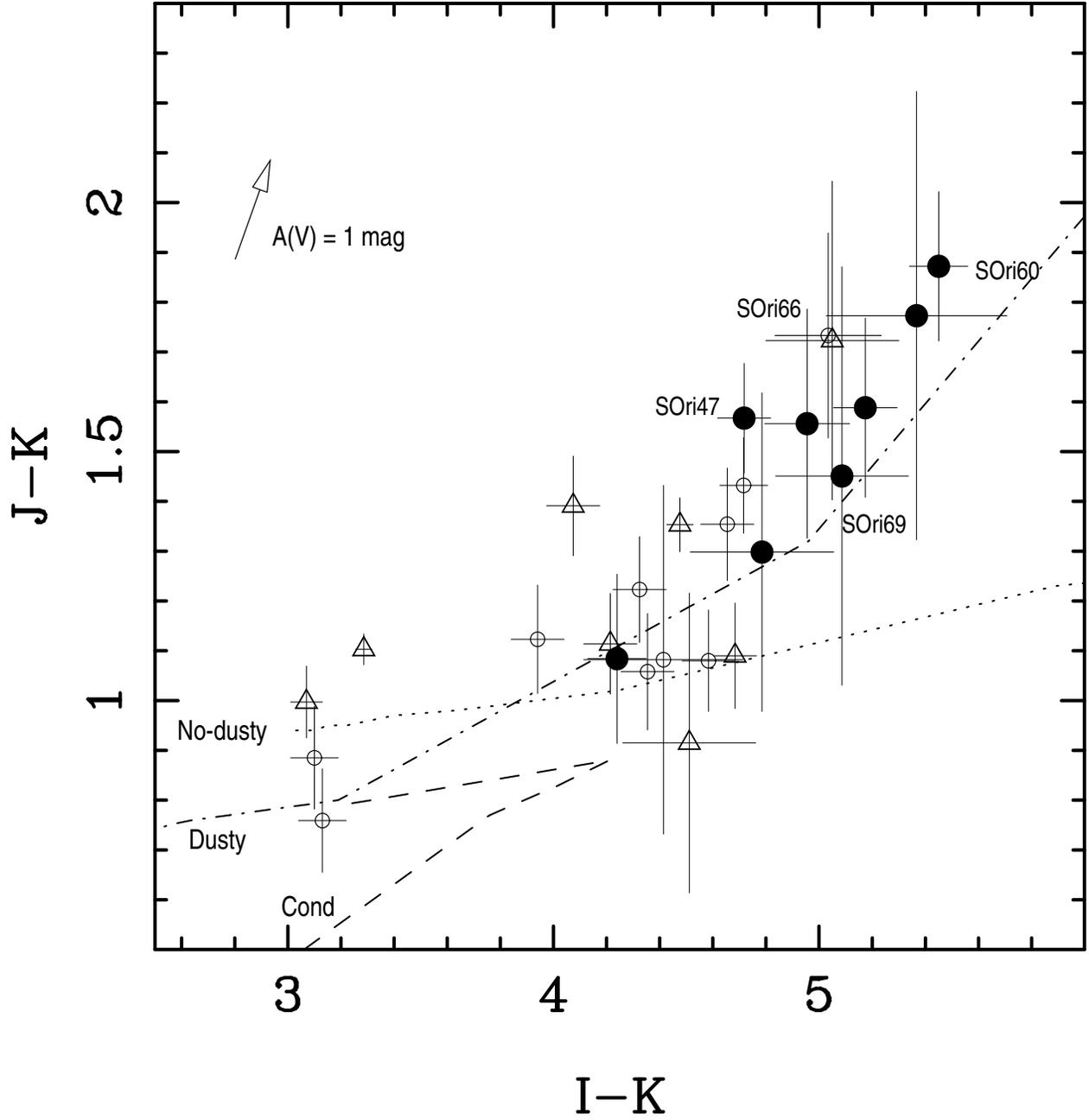}
\caption{\label{fig4} $J-K$ vs $I-K$
  color--magnitude diagram. The 5 Myr isochrones from Chabrier et
  al$.$ (\cite{chabrier00}) (Nextgen no-dusty models---dotted line,
  Dusty models---dot-dashed line, and Cond models---dashed line) are
  displayed.  Filled circles denote objects for which we have obtained
  $K$-band photometry with NIRC (Table~1).  Open symbols (triangles
  for objects without spectroscopic follow-up and circles for objects
  with spectroscopy) denote objects with photometry taken from B\'ejar
  et al$.$ (\cite{bejar01}) and Zapatero Osorio et al$.$
  (\cite{zapatero00}).}
\end{figure}

In Fig.~\ref{fig4}, we display a color-color diagram. Note that
S\,Ori\,66 and 69 are bluer than S\,Ori\,60, even though they are
fainter. Those relatively blue colors are probably due to partial
settling of dust grains below the photosphere. Fainter cluster members
are expected to have even bluer NIR colors. This should be taken into
account in future deep searches.


Our NIRC images have a scale of 0.15\arcsec/pix, and a total field of
view of 38\arcsec$\times$38\arcsec.  The FWHM of point sources in the
NIRC images of S\,Ori objects ranged from 2 to 4.5\,pix in the
$K$-band.  Only one of the possible cluster members appeared more
extended than the other sources in the same field. This object is
S\,Ori\,J053814.4--024012, which had a FWHM of 5.4\,pix (0.81\arcsec).
Its spectrum does not present water bands. This objects appears to be
an extremely red galaxy (ERG).  The NIR spectrum of ERGs is rather
featureless (Cimatti et al$.$ \cite{cimatti99}).

We have searched for candidate companions to the
\sigori~planetary-mass objects in the separation range 0.3\arcsec~to
10\arcsec, corresponding to projected semimajor axis, $a$, between
106\,AU and 3520\,AU at a distance of 352\,pc. Besides
S\,Ori\,J053814.4--024012, we find a resolved object at a separation
of 6.3\arcsec~north from S\,Ori\,66 with the following photometry:
$K$\,$\sim$\,19.5\,mag, $J-K$\,$\sim$\,0.9\,mag and
$I-K$\,$\sim$\,2.8\,mag. No other companion candidates are seen in our
NIRC images.  For the shorter separations, 0.3\arcsec~to 0.6\arcsec,
we are only sensitive to companions with nearly-equal mass as the
primary, but for the longer separations our 3\,$\sigma$ sensitivity
limit is K$\sim$19.5, which translates into a mass of only
0.005\,\msol~in \sigori. We note that our sensitivity limit for the
images of S\,Ori\,60 and 68 is brighter up to K$\sim$18.9\,mag.
Mart\'\i n et al$.$ (\cite{martin00}) have found a deficiency of
binaries with semimajor axis $\ge$27\,AU among Pleiades very low-mass
members.  Among field ultracool dwarfs, all known binaries with L-type
primaries have projected separations smaller than 10\,AU (Reid et
al$.$ \cite{reid01}). Clearly very low-mass binaries tend to be tight.
Is this a consequence of dynamical evolution (wide binaries are
loosely bound and can be disrupted) or formation (models of substellar
object formation via ejection in a multiple system predict a dearth of
wide substellar binaries, Reipurth \& Clarke \cite{reipurth01})?  The
lack of binaries with $a$\,$\ge$\,106\,AU among the substellar members
in \sigori~suggests that wide substellar binaries are also rare in a
very young cluster, favouring the idea that formation conditions may
be more important than dynamical evolution.  Our sample is, however,
too small to draw strong conclusions about the binary statistics of
the \sigori~substellar members.  Studies of future larger substellar
samples in \sigori~and other young associations will be necessary to
compare with the binary statistics of the Pleiades and other clusters
of similar age.

\acknowledgments Based on observations obtained at the W.\,M. Keck
Obs., which is operated as a scientific partnership among Caltech, UC
and NASA. The Observatory was made possible by the generous financial
support of the W.\,M. Keck Foundation. Partial financial support was
provided by the Spanish DGES PB98--0531--C02--02, by CICYT
ESP98--1339-CO2, and by NASA through a grant from STScI, which is
operated by the Association of Universities for Research in Astronomy,
Inc., under NASA contract NAS5-26555.

\clearpage
\begin{deluxetable}{lccccc}
\scriptsize
\tablecaption{\label{tab1} 
  Data for \sigori~candidate members with planetary masses}
\tablewidth{0pt}
\tablehead{
 \colhead{Name} & \colhead{$I$} & \colhead{$K$} &  \colhead{$I-K$}  & 
 \colhead{$J-K$} & Sp. Type }
\startdata
S\,Ori\,47 & 20.53$\pm$0.05 & 15.81$\pm$0.08 & 4.72$\pm$0.10 &
1.57$\pm$0.11 & L1.5$\pm$0.5 \nl S\,Ori\,J053814.4--024012 &
24.2$\pm$0.4 & 19.02$\pm$0.15 & 5.2$\pm$0.5 & 1.5$\pm$0.7 & ERG \nl
S\,Ori\,57 & 21.87$\pm$0.02 & 18.12$\pm$0.15 & 3.75$\pm$0.15 &
0.51$\pm$0.16 & NM \nl S\,Ori\,60 & 22.75$\pm$0.05 & 17.30$\pm$0.10 &
5.45$\pm$0.11 & 1.87$\pm$0.15 & L5.5$\pm$1.0 \nl S\,Ori\,62 &
23.03$\pm$0.06 & 17.86$\pm$0.10 & 5.17$\pm$0.12 & 1.59$\pm$0.18 &
L4.0$\pm$1.0 \nl S\,Ori\,66 & 23.23$\pm$0.12 & 18.27$\pm$0.10 &
4.96$\pm$0.16 & 1.56$\pm$0.23 & L6.0$\pm$1.0 \nl S\,Ori\,67 &
23.40$\pm$0.09 & 18.62$\pm$0.25 & 4.78$\pm$0.27 & 1.30$\pm$0.32 &
L5.0$\pm$2.0 \nl S\,Ori\,68 & 23.77$\pm$0.17 & 18.41$\pm$0.30 &
5.37$\pm$0.34 & 1.77$\pm$0.45 & L5.0$\pm$2.0 \nl S\,Ori\,69 &
23.89$\pm$0.15 & 18.81$\pm$0.20 & 5.09$\pm$0.25 & 1.45$\pm$0.42 & T0:
\nl
\enddata
\tablecomments{$I$-band and $J$-band data have been previously
  presented by Zapatero Osorio et al$.$ (\cite{zapatero00}). ERG:
  Extremely red galaxy. NM: non-member of the cluster.  Spectral types
  for S\,Ori\,47 and 62 were taken from Zapatero-Osorio et al$.$
  (\cite{zapatero00}).  Spectral types for S\,Ori\,67 and 68 were
  taken from Barrado et al$.$ (\cite{barrado01}).  A colon in the
  classification of S\,Ori\,69 indicates that the spectral type is
  tentative. All spectral types are based on optical data except for
  S\,Ori\,60, 62, 66 and 69, which are based on NIR spectra.}
\end{deluxetable}

\clearpage

\clearpage

\clearpage

\clearpage

\end{document}